\documentclass[journal]{IEEEtran}
\ifCLASSINFOpdf
\else
\fi
\usepackage{amsmath,amssymb,amsthm}

\usepackage{authblk}

\usepackage{hhline}
\usepackage{graphicx}
\usepackage{epstopdf}
\usepackage{subcaption}
\usepackage{algorithm}
\usepackage{algorithmic}
\usepackage{amsmath}
\usepackage{cite}

\begin{document}
	\title{Efficient 3D
		Placement of a UAV Using Particle Swarm Optimization}
	\author[1]{Hazim Shakhatreh}
	\author[1]{Abdallah Khreishah}
	\author[2]{Ayoub Alsarhan}
	\author[3]{Issa Khalil}
	\author[4]{Ahmad Sawalmeh}
	\author[4]{Noor Shamsiah Othman}

	\affil[1]{Department of Electrical and Computer Engineering, New Jersey Institute of Technology}
	\affil[2]{Department of Computer Information System, Hashemite University}
	\affil[3]{Qatar Computing Research Institute, Hamad bin Khalifa University}
	\affil[4]{UNITEN, Selangor, Malaysia}
	
	\maketitle
	
	\begin{abstract}
		Unmanned aerial vehicles (UAVs) can be used as aerial wireless base stations when cellular networks go down. Prior studies on UAV-based wireless coverage typically consider an Air-to-Ground path loss model, which assumes that the users are outdoor and they are located on a 2D plane. In this paper, we propose using a single UAV to provide wireless coverage for indoor users inside a high-rise building under disaster situations (such as earthquakes or floods), when cellular networks are down. We assume that the locations of indoor users are uniformly distributed in each floor and we propose a particle swarm optimization algorithm to find an efficient 3D placement of a UAV that minimizes the total transmit power required to cover the indoor users. 
	\end{abstract}
	
	
	\section{Introduction}
	\label{sec:Introduction}
	UAVs can be used to provide wireless coverage during emergency cases where each UAV serves as an aerial wireless base station when the cellular network goes down~\cite{bupe2015relief}. They can also be used to supplement the ground base station in order to provide better coverage and higher data rates for the users~\cite{bor2016efficient}.
	
	In order to use a UAV as an aerial wireless base station, the authors in~\cite{al2014modeling} presented an Air-to-Ground path loss model that helped the academic researchers to formulate many important problems. The authors of~\cite{mozaffari2015drone,mozaffari2016optimal,mozaffari2016efficient,kalantari2016number,shakhatreh2016continuous} utilized this model to study the problem of UAV placement, where the objective is to minimize the number of UAVs for covering a given area.
	The authors of~\cite{mozaffari2015drone} described the tradeoff in this model. At a low altitude, the path loss between the UAV and the ground user decreases, while the probability of line of sight links also decreases. On the other hand, at a high altitude line of sight connections exist with a high probability, while the path loss increases.
	However, it is assumed that all users are outdoor and the location of each user can be represented by an outdoor 2D point. These assumptions limit the applicability of this model when one needs to consider indoor users.
	
	Providing good wireless coverage for indoor users is very important. According to Ericsson report~\cite{ericsson}, 90\% of the time people are indoor and 80\% of the mobile Internet access traffic also happens indoors~\cite{alcatel,cisco}. To guarantee the wireless coverage, the service providers are faced with several key challenges, including providing service to a large number of indoor users and the ping pong effect due to interference from near-by macro cells~\cite{comm,amplitic,zhang2016study}. In this paper, we propose using a single UAV to provide wireless coverage for users inside a high-rise building during emergency cases, when the cellular network service is not available.
		\begin{figure}[!t]
			\centering
			\includegraphics[scale=0.5]{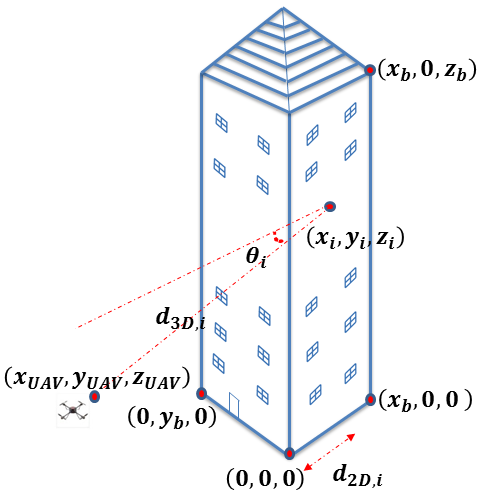}
			\caption{Parameters of the            
				path loss model}
			\label{fig:figure1}
		\end{figure}

	In~\cite{haz2017efficient}, we study the problem of efficient UAV placement, where the objective is to
	minimize the total transmit power required to cover the entire high-rise building. We consider two cases
	of practical interest and provide efficient solutions to the formulated problem under these cases. In the first case, we find the minimum transmit power such that an indoor user with the maximum path loss can be covered. In the second case, we assume that the locations of indoor users are symmetric across the dimensions of each floor and we propose a gradient descent algorithm to find an efficient 3D placement of a UAV. Our main contribution in this paper is to study the problem of efficient UAV placement, where the objective is to minimize the total transmit power required to cover the entire high-rise building, when the locations of indoor users are uniformly distributed in each floor, we propose a particle swarm optimization algorithm for finding an efficient location of the UAV.

	The rest of this paper is organized as follows. In Section II, we describe the system model and a path loss model suitable for studying indoor wireless coverage. In Section III, we formulate the problem of UAV placement with an objective of minimizing the transmit power for covering the entire building. In Section IV, we present the particle swarm optimization algorithm and show how to find an efficient placement of the UAV such that the total transmit power is minimized. Finally, we present our numerical results in Section V and make concluding remarks in Section VI.
	
	\section{System Model}
	\label{sec:system_model}
	\subsection{System Settings}
	\label{subsec:system settings}
	Let ($x_{UAV}$,$y_{UAV}$,$z_{UAV}$) denote the 3D location of the UAV. We assume that all users are located inside a high-rise building as shown in Figure~\ref{fig:figure1}, and use ($x_{i}$,$y_{i}$,$z_{i}$) to denote the location of user $i$. The dimensions of the high-rise building are $[0,x_b]$ $\times$ $[0,y_b]$ $\times$ $[0,z_b]$. Also, let $d_{3D,i}$ be the 3D distance between the UAV and indoor user $i$, let $\theta_{i}$ be the incident angle , and let $d_{2D,i}$ be the 2D indoor distance of user $i$ inside the building.
	
	\subsection{Outdoor-Indoor Path Loss Model}
	\label{subsec:Outdoor-Indoor Path Loss Model}
	The Air-to-Ground path loss model presented in~\cite{al2014modeling} is not appropriate when we consider wireless coverage for indoor users, because this model assumes that all users are outdoor and located at 2D points. In this paper, we adopt the Outdoor-Indoor path loss model, certified by the ITU~\cite{series2009guidelines}. The path loss is given as follows:\\
	\begin{equation}
	\begin{split}
	L_i=L_{F}+L_{B}+L_{I}= ~~~~~~~~~~~~~~~~\\
	(w\log_{10}d_{3D,i}+w\log_{10}f_{Ghz}+g_{1})+\\
	(g_{2}+g_{3}(1-\cos\theta_{i})^{2})+(g_{4}d_{2D,i})~~~~
	\end{split}
	\end{equation}
	where $L_{F}$ is the free space path loss, $L_{B}$ is the building penetration loss, and $L_{I}$ is the indoor loss. In this model, we also have
	$w$=20, $g_{1}$=32.4, $g_{2}$=14, $g_{3}$=15, $g_{4}$=0.5~\cite{series2009guidelines} and $f_{Ghz}$ is the carrier frequency (2Ghz). Note that there is a key tradeoff in the above model when the horizontal distance between the UAV and a user changes. When this horizontal distance increases, the free space path loss (i.e., $L_F$) increases as $d_{3D,i}$ increases, while the building penetration loss (i.e., $L_B$) decreases as the incident angle (i.e., $\theta_i$) decreases. Similarly, when this horizontal distance decreases, the free space path loss (i.e., $L_F$) decreases as $d_{3D,i}$ decreases, while the building penetration loss (i.e., $L_B$) increases as the incident angle (i.e., $\theta_i$) increases. 
	\section{Problem Formulation }
	Consider a transmission between a UAV located at ($x_{UAV}$,$y_{UAV}$,$z_{UAV}$) and an indoor user $i$ located at ($x_i$,$y_i$,$z_i$). The rate for user $i$ is given by:
	\begin{equation}
	\begin{split}
			C_{i}=Blog_{2}(1+\dfrac{P_{t,i}/L_i}{N})
	\end{split}
	\end{equation}
	where $B$ is the transmission bandwidth of the UAV, $P_{t,i}$ is the UAV transmit power to indoor user $i$, $L_i$ is the path loss between the UAV and indoor user $i$ and $N$ is the noise power. In this paper, we do not explicitly model interference, and instead, implicitly model it as noise.
	
	Let us assume that each indoor user has a channel with bandwidth equals $B/M$, where $M$ is the number of users inside the building and the rate requirement for each user is $v$. Then the minimum power required to satisfy this rate for each user is given by:
	 \begin{equation}
	 \begin{split}
	 P_{t,i,min}=(2^{\frac{v.M}{B}}-1)\star N\star L_i
	 \end{split}
	 \end{equation}
	 Our goal is to find an efficient location of the UAV such that the total transmit power required to satisfy the rate requirement of each indoor user is minimized. The objective function can be represented as:
	 \begin{equation}
	 \begin{split}
	 P=\sum_{i=1}^{M}(2^{\frac{v.M}{B}}-1)\star N\star L_i,\\
	 \end{split}
	 \end{equation}
	 where $P$ is the UAV total transmit power. Since $(2^{\frac{v.M}{B}}-1)\star N$ is constant, our problem can be formulated as:
	  \begin{equation}
	  \begin{split}
	  \min_{x_{UAV},y_{UAV},z_{UAV}} L_{Total}=\sum_{i=1}^{M}L_i~~~~~~~~~~~~~~~~~~~~~~~~~~~~~~~~~\\
	  subject ~to~~~~~~~~~~~~~~~~~~~~~~~~~~~~~~~~~~~~~~~~~~~~~~~~~~~~~~~~~~~\\
	  x_{min}\leq x_{UAV}\leq x_{max},~~~~~~~~~~~~~~~~~~~~~~\\
	  y_{min}\leq y_{UAV}\leq y_{max},~~~~~~~~~~~~~~~~~~~~~~\\
	  z_{min}\leq z_{UAV}\leq z_{max},~~~~~~~~~~~~~~~~~~~~~~\\
	  L_{Total}\leq L_{max}~~~~~~~~~~~~~~~~~~~~~~~~~~~
	   \end{split}
	   \end{equation}
	   Here, the first three constraints represent the minimum and maximum allowed values for $x_{UAV}$, $y_{UAV}$ and  $z_{UAV}$. In the fourth constraint, $L_{max}$ is the maximum allowable path loss and equals $P_{t,max}$$/$$((2^{\frac{v.M}{B}}-1)\star N)$, where $P_{t,max}$ is the maximum transmit power of UAV.
	
	   Finding the optimal placement of UAV is generally  difficult because the problem is non-convex. Therefore, in the next section, we present the particle swarm optimization to find an efficient solution for the formulated problem.
	    
	   \section{Efficient Placement of UAV}
	   \label{sec:optimal}
	   Due to the intractability of the problem, we propose the Particle Swarm Optimization (PSO)~\cite{kennedy1995particle,clerc2002particle} to find an efficient 3D placement of the UAV, when the locations of indoor users are uniformly distributed in each floor. In~\cite{haz2017efficient}, we prove that $z_{UAV}$=$0.5z_{b}$ and $y_{UAV}$=$0.5y_{b}$ when the locations of indoor users are symmetric across the dimensions of each floor. Then, we  use the gradient descent algorithm to find $x_{UAV}$ that minimizes the transmit power required to cover the building. 
	   
	   The particle swarm optimization algorithm starts with (npop) random solutions and iteratively tries to improve the candidate solutions based on the best experience of each candidate (particle(i).best.location) and the best global experience (globalbest.location). In each iteration, the best location for each particle (particle(i).best.location) and the best global location (globalbest.location) are updated and the velocities and locations of the particles are calculated based on them~\cite{kalantari2016number}. The velocity is given by:
	   \begin{equation}
	   \begin{split}
	   particle(i).velocity=w*particle(i).velocity+\\c1*rand(varsize).*(particle(i).best.location\\
	   -particle(i).location)+c2*rand(varsize).*\\
	   (globalbest.location-particle(i).location)
	   \end{split}
	   \end{equation}
	   where $w$ is the inertia weight, $c1$ and $c2$ are the personal and global learning coefficients, and $rand(varsize)$ are random positive numbers. Also, the location of each particle is updated as:
	    \begin{equation}
	    \begin{split}  
	   particle(i).location=particle(i).location\\
	   +particle(i).velocity
	       \end{split}
	       \end{equation}

	       The pseudo code of the PSO algorithm is shown in Algorithm 1. The number of iterations (maxit) in the algorithm should be high enough to guarantee the stability.
	   	\begin{algorithm}
	   		\begin{algorithmic}
	   			\STATE \textbf{Input:}
	   			\STATE The lower and upper bounds of decision variable (varmin,varmax), Construction coefficients ($\kappa$,$\phi_1$,$\phi_2$), Maximum number of iterations (maxit), Population size (npop)
	   			\STATE \textbf{Initialiaztion:}
	   			\STATE $\phi$=$\phi_1$+$\phi_1$, $\chi$ = ${2\kappa}/{|2-\phi-(\phi^2-4\phi)^{0.5}|}$
	   			\STATE w=$\chi$, c1=$\chi$$\phi_1$, c2=$\chi$$\phi_2$, globalbest.cost=inf
	   			\STATE \textbf{for} i=1:npop
	   			\STATE ~~~~~particle(i).location=unifrnd(varmin, varmax, varsize)
	   			\STATE~~~~~particle(i).velocity=zeros(varsize)
	   			\STATE~~~~~particle(i).cost=costfunction(particle(i).location)
	   			\STATE~~~~~particle(i).best.location=particle(i).location
	   			\STATE~~~~~particle(i).best.cost=particle(i).cost
	   			\STATE~~~~~\textbf{if} particle(i).best.cost $<$ globalbest.cost
	   			\STATE~~~~~~~~globalbest=particle(i).best
	   			\STATE~~~~~\textbf{end if}
	   			\STATE  \textbf{end}
	   			\STATE \textbf{PSO Loop:}
	   			\STATE \textbf{for} t=1:maxit
	   			\STATE~~~~~\textbf{for }i=1:npop
	   			\STATE~~~~~~~~particle(i).velocity=w*particle(i).velocity+
	   			\STATE ~~~~~~~~c1*rand(varsize).*(particle(i).best.location-
	   			\STATE ~~~~~~~~particle(i).location)+c2*rand(varsize).*
	   			\STATE ~~~~~~~~(globalbest.location-particle(i).location)
	   			\STATE~~~~~~~~particle(i).location=particle(i).location+
	   			\STATE ~~~~~~~~~~~~~~~~~~~~~~~~~~~~~~~particle(i).velocity
	   			\STATE~~~~~~~~particle(i).cost=costfunction(particle(i).location)
	   			\STATE~~~~~~~~~~ \textbf{if} particle(i).cost $<$ particle(i).best.cost
	   			\STATE~~~~~~~~~~~~~~particle(i).best.location = particle(i).location
	   			\STATE~~~~~~~~~~~~~~particle(i).best.cost = particle(i).cost
	   			\STATE~~~~~~~~~~~~~~~~~~\textbf{if} particle(i).best.cost $<$ globalbest.cost
	   			\STATE~~~~~~~~~~~~~~~~~~~~~~globalbest=particle(i).best
	   			\STATE~~~~~~~~~~~~~~~~~~\textbf{end if}
	   			\STATE~~~~~~~~~~ \textbf{end if}
	   			\STATE ~~~~~\textbf{end}
	   			\STATE\textbf{end}
	   			
	   		\end{algorithmic}
	   		\caption{Efficient UAV placement using PSO algorithm}
	   	\end{algorithm}

	    \section{Numerical Results}
	    \label{sec:results}
	    First, we assume that each floor contains 20 users and the locations of indoor users are symmetric across the dimensions of each floor. Then, we apply the particle swarm optimization algorithm to find an efficient 3D placement of a UAV. Table I lists the parameters used in the numerical analysis. The particle swarm optimization algorithm will converge to the efficient 3D UAV placement when the maximum number of iterations is equal to 50. On the other hand, the gradient descent algorithm will converge to the efficient placement when the maximum number of iterations is equal to 100 and the step tolerance is equal to 0.01.  
	     \begin{table}[!h]
	     	\scriptsize
	     	\renewcommand{\arraystretch}{1.3}
	     	\caption{Parameters in numerical analysis}
	     	\label{table}
	     	\centering
	     	\begin{tabular}{|c|c|}
	     		\hline
	     		Hight of each floor & 5 meters\\
	     		\hline
	     		Population size (npop) & 50\\
	     		\hline
	     		Maximum number of iterations (maxit) & 50\\
	     		\hline 
	     		The carrier frequency $f_{Ghz}$ & 2Ghz\\
	     		\hline 
	     		Number of users in each floor & 20 users\\
	     		\hline
	     		(varmin,varmax) & (-1000,1000)\\
	     		\hline
	     		varsize & 3\\
	     		\hline
	     		($\kappa$,$\phi_1$,$\phi_2$) & (1,2.05,2.05)\\
	     		\hline
	     	\end{tabular}
	     \end{table}
	     \begin{table*}[!t]
	     	\scriptsize
	     	\renewcommand{\arraystretch}{1.55}
	     	\caption{Simulation Results}
	     	\label{table}
	     	\centering
	     	\begin{tabular}{|c||c|c|c|c|c|c|}
	     		\hline \hline
	     		Algorithm&Distribution (locations of indoor users) &Building height &Horizontal building&Vertical building &Efficient 3D placement &Efficient total \\
	     		&&$z_b$& width $x_b$&width $y_b$&&path loss(dB) \\
	     		\hline \hline
	     		GD&symmetric across the dimensions of each floor&200&20&50&(-24.7967, 25, 100)&$7.6733*10^{4}$ \\
	     		\hline
	     		PSO &symmetric across the dimensions of each floor&200&20&50&(-24.7491, 24.9419, 100.0491)&$7.6733*10^{4}$\\
	     		\hline \hline
	     		GD&symmetric across the dimensions of each floor&250&20&50&(-35.2978, 25, 125)&$9.7381*10^{4}$ \\
	     		\hline
	     		PSO &symmetric across the dimensions of each floor&250&20&50&(-35.3077, 24.9162, 125.0544)&$9.7381*10^{4}$\\
	     		\hline \hline
	     		GD&symmetric across the dimensions of each floor&300&20&50&(-45.1131, 25, 150)&$1.1837*10^{5}$ \\
	     		\hline
	     		PSO &symmetric across the dimensions of each floor&300&20&50&(-45.1352, 25.0371, 149.7681)&$1.1837*10^{5}$\\
	     		\hline \hline
	     		
	     		\hline \hline
	     		GD&uniformly distributed in each floor&200&20&50&(-24.7254, 25, 100)&$7.8853*10^{4}$ \\
	     		\hline
	     		PSO &uniformly distributed in each floor&200&20&50&(-21.7995, 37.3891, 111.7901)&$7.8645*10^{4}$\\
	     		\hline \hline
	     		GD&uniformly distributed in each floor&250&20&50&(-33.8180, 25, 125)&$9.9855*10^{4}$ \\
	     		\hline
	     		PSO &uniformly distributed in each floor&250&20&50&(-32.9212, 28.7125, 124.0291)&$9.9725*10^{4}$\\
	     		\hline \hline
	     		GD&uniformly distributed in each floor&300&20&50&(-43.1170, 25, 150)&$1.2154*10^{5}$ \\
	     		\hline
	     		PSO &uniformly distributed in each floor&300&20&50&(-46.5898, 31.5061 ,143.8588)&$1.2117*10^{5}$\\
	     		\hline \hline

	     		\hline \hline
	     		GD&uniformly distributed in each floor&250&10&50&(-38.5210, 25, 125)&$9.7413*10^{4}$ \\
	     		\hline
	     		PSO &uniformly distributed in each floor&250&10&50&(-32.1042, 21.0174, 129.2663)&$9.7252*10^{4}$\\
	     		\hline \hline
	     		GD&uniformly distributed in each floor&250&30&50&(-29.3930, 25, 125)&$1.0275*10^{5}$ \\
	     		\hline
	     		PSO &uniformly distributed in each floor&250&30&50&(-25.5294, 4.9387, 138.7650)&$1.0211*10^{5}$\\
	     		\hline \hline
	     		GD&uniformly distributed in each floor&250&50&50&(-22.7119, 25, 125)&$1.0753*10^{5}$ \\
	     		\hline
	     		PSO &uniformly distributed in each floor&250&50&50&(-14.5488 17.3082 131.8940)&$1.0696*10^{5}$\\
	     		\hline \hline
	     		
	     	\end{tabular}
	     \end{table*}

	      \begin{figure}[!h]
	      	\centering
	      	\includegraphics[scale=0.21]{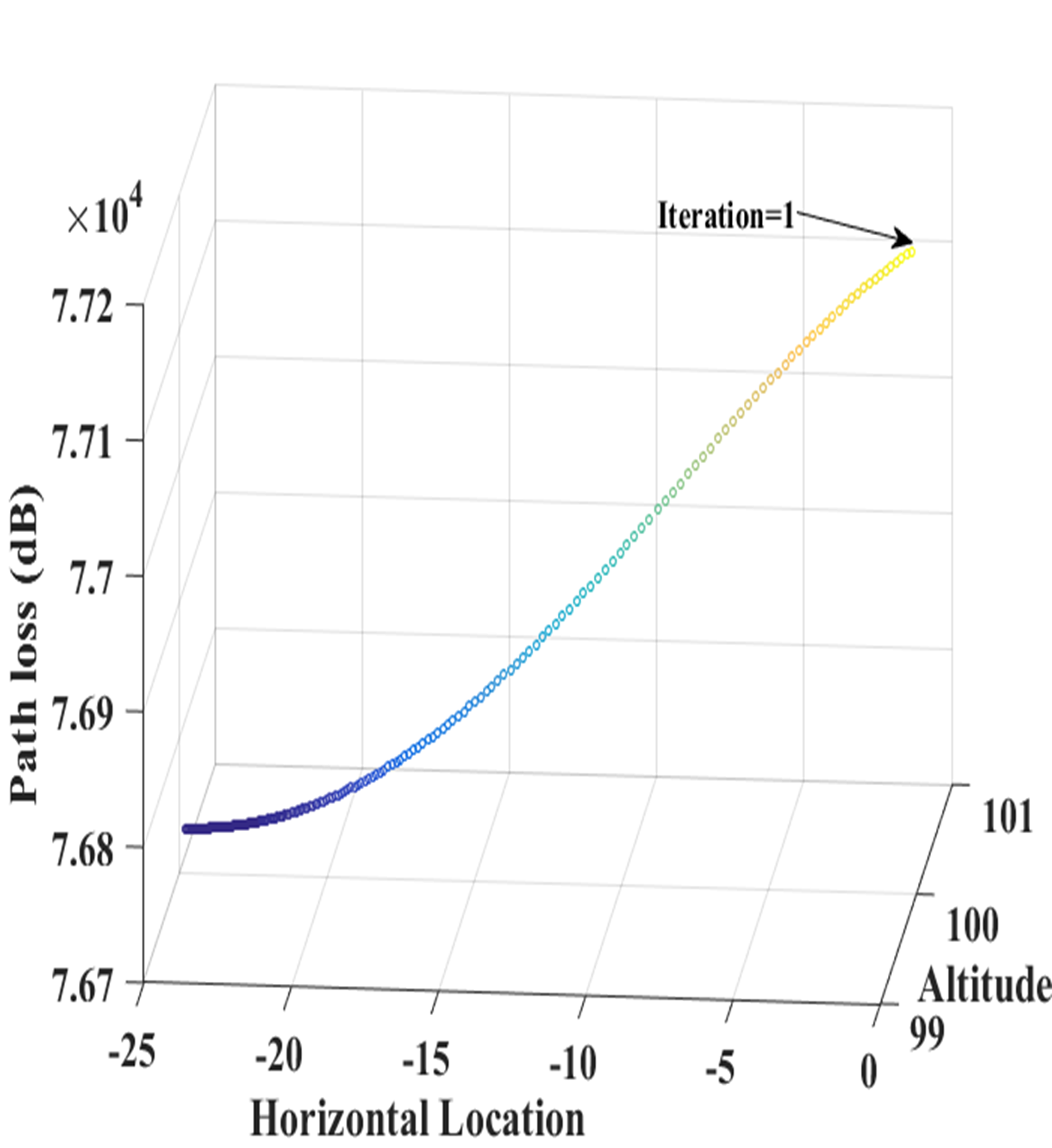}
	      	\caption{The convergence speed of the GD algorithm for 200 meters building height}
	      	\label{fig:figuretwo}
	      \end{figure}

	      \begin{figure}[!h]
	      	\centering
	      	\includegraphics[scale=0.21]{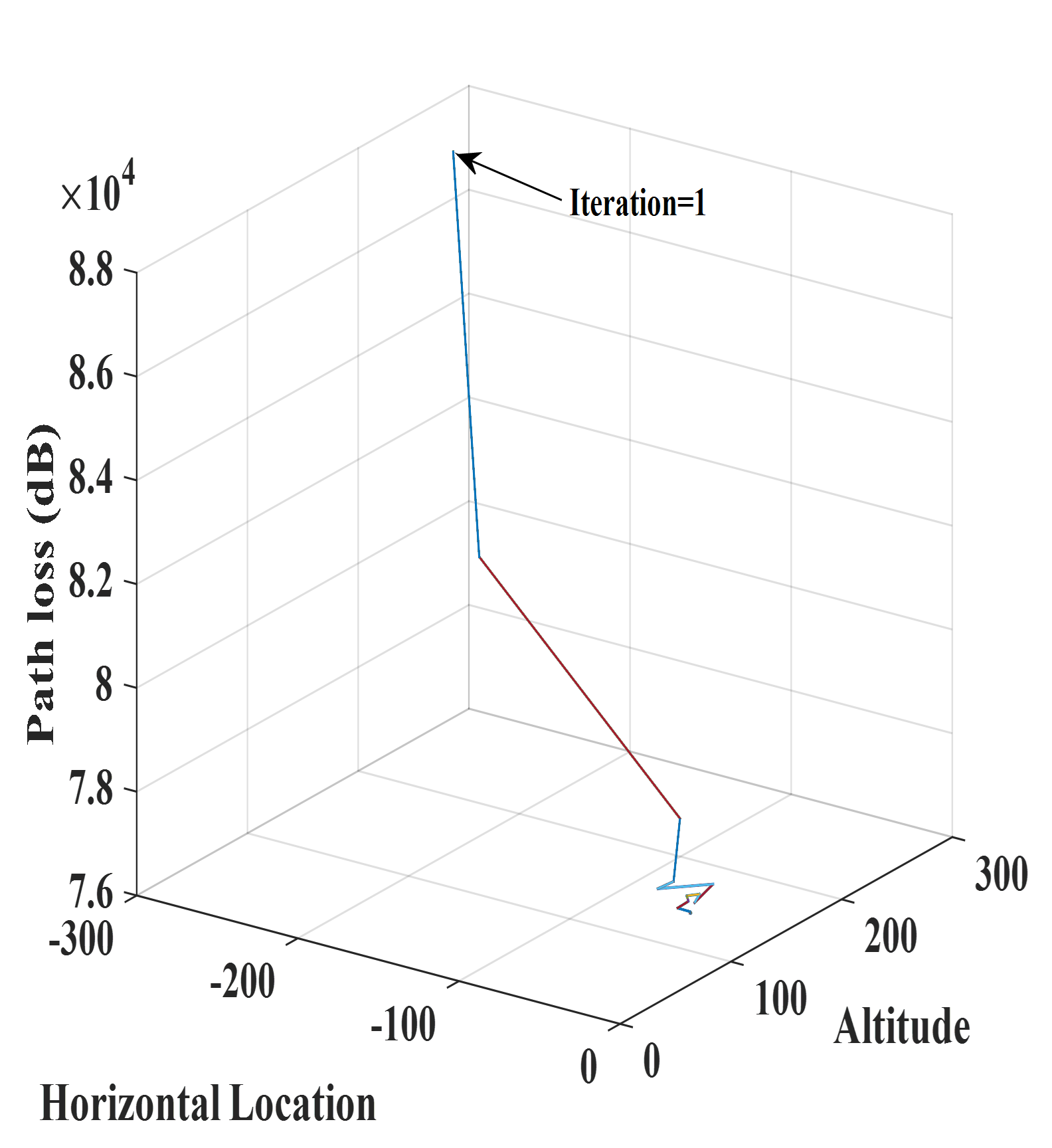}
	      	\caption{The convergence speed of the PSO algorithm for 200 meters building height}
	      	\label{fig:figurethree}
	      \end{figure}

	    In Figure~\ref{fig:figuretwo}, we show the convergence speed of the gradient descent algorithm when the building hight is 200 meters. The 3D efficient placement is (-24.7967, 25, 100) and the total path loss is $7.6733*10^{4}$. The convergence speed of the particle swarm optimization algorithm when the building hight is 200 meters is shown in Figure~\ref{fig:figurethree}. The 3D efficient placement is (-24.7491, 24.9419, 100.0491) and the total path loss is ($7.6733*10^{4}$). Table II lists the simulation results for 250 meters and 300 meters building heights. As can be seen from the simulation results, both of the algorithms converge to the same 3D placement. 
	     
	     After that, we assume that each floor contains 20 users and the locations of these users are uniformly distributed in each floor.

	     \begin{figure*}[!t]
	     	\centering
	     	\includegraphics[scale=0.19]{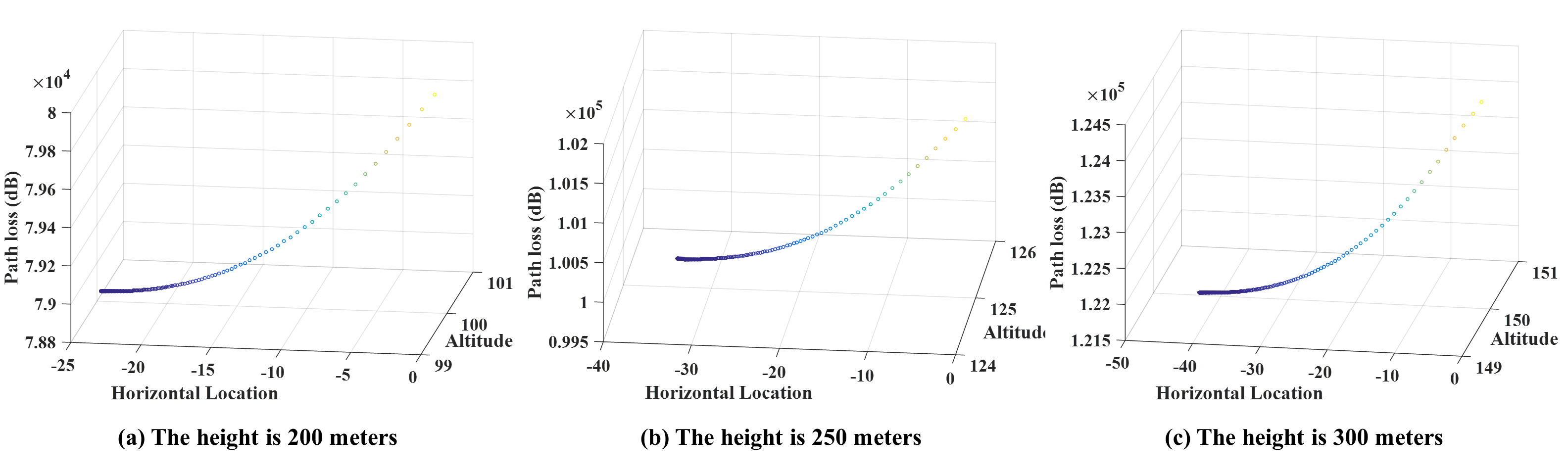}
	     	\caption{The convergence speeds of the GD algorithm for different building heights}
	     	\label{fig:figurefour}
	     \end{figure*}
	     \begin{figure*}[!t]
	     	\centering
	     	\includegraphics[scale=0.19]{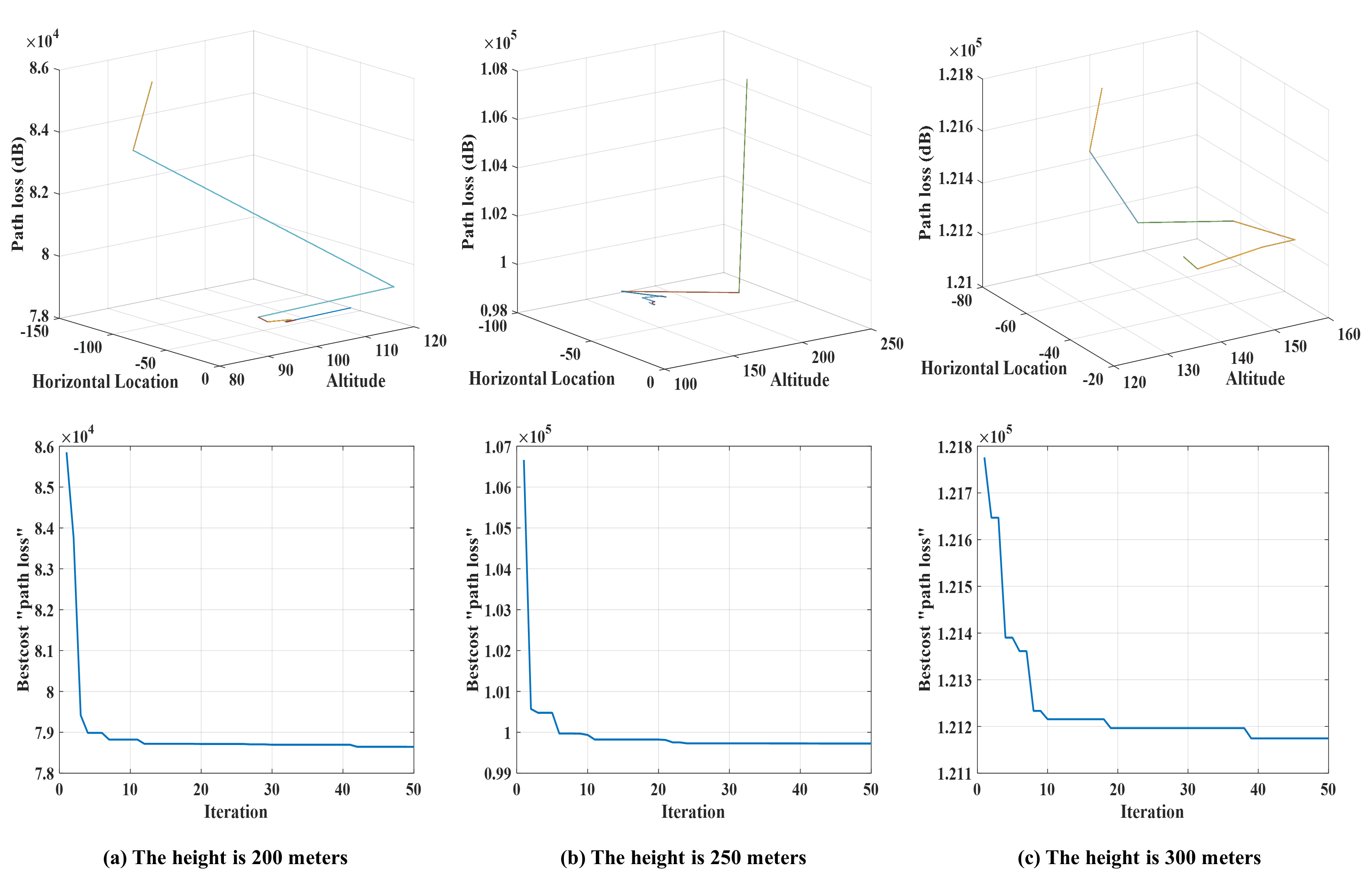}
	     	\caption{The convergence speeds of the PSO algorithm for different building heights}
	     	\label{fig:figurefive}
	     \end{figure*}
	     
	      In Figure~\ref{fig:figurefour},  we show the convergence speeds of the gradient descent algorithm for different building heights. The 3D efficient placements and the total costs for 200 meter, 250 meter and 300 meter buildings are (-24.7254, 25, 100) ($7.8853*10^{4}$), (-33.8180, 25, 125) ($9.9855*10^{4}$) and (-43.1170, 25, 150)($1.2154*10^{5}$), respectively. The convergence speeds of the particle swarm optimization algorithm for different building heights are shown in Figure~\ref{fig:figurefive}. The 3D efficient placements and the total costs for 200 meter, 250 meter and 300 meter buildings are (-21.7995, 37.3891, 111.7901) ($7.8645*10^{4}$), (-32.9212, 28.7125, 124.0291) ($9.9725*10^{4}$) and (-46.5898, 31.5061 ,143.8588)($1.2117*10^{5}$), respectively. As can be seen from the simulation results, the PSO algorithm provides better results, it provides total cost less than the cost that the GD algorithm provides by (37dB-208dB). This is because the PSO algorithm is designed for the case in which the locations of indoor users are uniformly distributed in each floor. On the other hand, the GD algorithm is designed for the case in which the locations of indoor users are symmetric across the dimensions of each floor.
	      
\begin{figure*}[!t]
	      	\centering
	      	\includegraphics[scale=0.19]{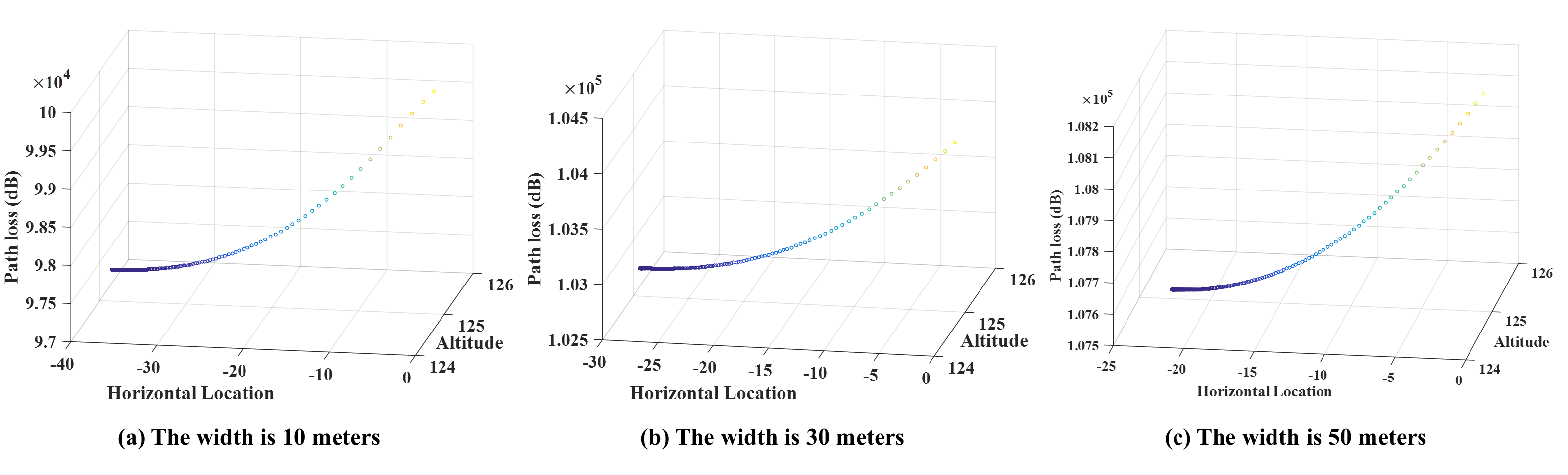}
	      	\caption{The convergence speeds of the GD algorithm for different building widths}
	      	\label{fig:figuresix}
	      \end{figure*}

\begin{figure*}[!t]
	      	\centering
	      	\includegraphics[scale=0.19]{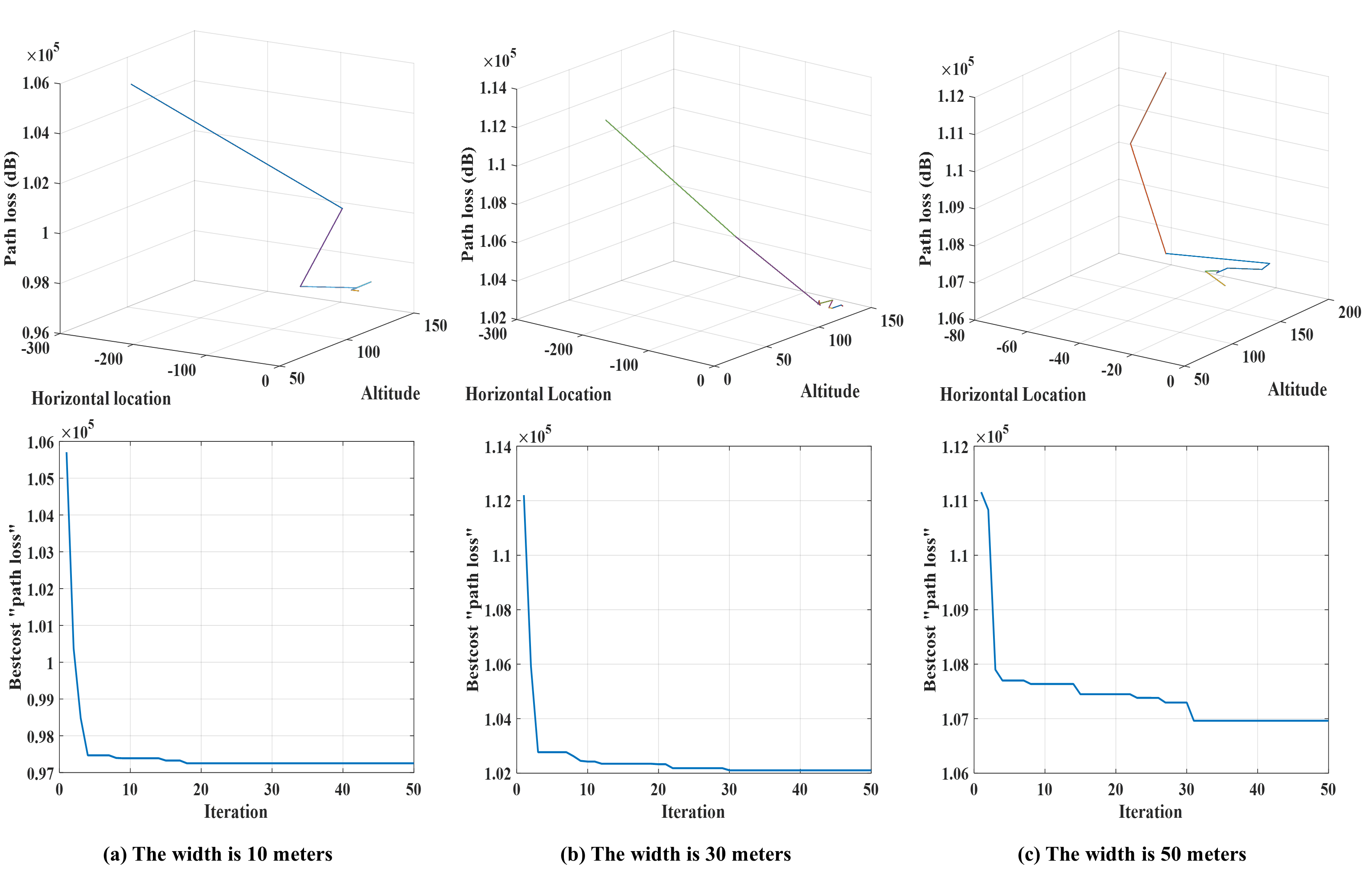}
	      	\caption{The convergence speeds of the PSO algorithm for different building widths}
	      	\label{fig:figureseven}
	      \end{figure*}

	      We investigate the impact of different building widths (i.e., $x_b$) in Figures ~\ref{fig:figuresix} and~\ref{fig:figureseven} using the GD and PSO algorithms. We fix the building height to be 250 meters and vary the building width.  As can be seen from the simulation results, the PSO algorithm provides better results, it provides total cost less than the cost that the GD algorithm provides by (57dB-161dB).
	      
	       Table II lists the simulation results. We can notice that when the height of the building increases, the efficient horizontal point $x_{UAV}$ increases. This is to compensate the increased building penetration loss due to an increased incident angle. Also, when the building width increases, the efficient horizontal distance decreases. This is to compensate the increased indoor path loss due to an increased building width.
	       

	    \section{Conclusion}
	    \label{sec:Conclusion}
	    In this paper, we study the problem of providing wireless coverage for users inside a high-rise building using a single UAV. Due to the intractability of the problem, we propose the particle swarm optimization algorithm to find an efficient 3D placement of a UAV that minimizes the total transmit power required to cover the indoor users when the same number of users is uniformly distributed in each floor. In order to model more realistic scenarios, we will consider different types of user distribution in our future work. We will also study the problem of providing wireless coverage using multiple UAVs. 
	    \section*{Acknowledgment}
	    This work was supported in part by the NSF under Grant CNS-1647170.
	\bibliographystyle{IEEEtran}
	\bibliography{UAVpath}

\end{document}